\documentclass[magazine]{IEEEtran}

\usepackage[T1]{fontenc}
\usepackage[utf8]{inputenc}
\usepackage{cite}
\usepackage{url}
\usepackage{amsmath, mathtools}
\usepackage{amsfonts,amsthm,bm}
\usepackage{amssymb}
\usepackage{comment}
\usepackage{graphicx}
\usepackage{standalone}
\usepackage{dsfont}
\usepackage{mathtools}
\usepackage{color, colortbl}
\usepackage{soul}
\usepackage[normalem]{ulem}
\usepackage[acronym,shortcuts]{glossaries}
\usepackage{arydshln}
\usepackage{bbm}
\usepackage{svg} 
\usepackage{algorithm,algorithmic}
\usepackage{adjustbox}
\usepackage[inline]{enumitem}
\usepackage{multirow}
\usepackage{tabularx}
\usepackage{booktabs}

\usepackage{comment}
\usepackage{tikz}
\usepackage{pgfplots}
\pgfplotsset{compat=newest}
\pgfplotsset{plot coordinates/math parser=false}
\newlength\fheight
\newlength\fwidth
\usetikzlibrary{plotmarks,patterns,decorations.pathreplacing,backgrounds,calc,arrows,arrows.meta,spy,matrix}
\usepgfplotslibrary{patchplots,groupplots}
\usepackage{tikzscale}
\usepackage{balance}
\usepackage{stfloats} 

\usepackage[capitalize]{cleveref}
\crefname{section}{Sec.}{Secs.}

\usepackage{subcaption}
\usepackage{graphicx}
\usepackage{caption}
\usepackage{xcolor}
\definecolor{ieeeblue}{HTML}{00629B}
\definecolor{ieeeorange}{HTML}{FFA300}
\definecolor{ieeegreen}{HTML}{78BE20}
\definecolor{ieeered}{HTML}{BA0C2F}

\usetikzlibrary {patterns.meta}

\usetikzlibrary{patterns}

\usepackage{ulem}

\providecolor{added}{rgb}{0,0,1}
\providecolor{deleted}{rgb}{1,0,0}

\makeatletter
\newcommand{\new}[1]{{\textcolor{blue}{#1}}}

\makeatletter
\newcommand\remembertext[2]{
  \immediate\write\@auxout{\unexpanded{\global\long\@namedef{mytext@#1}{#2}}}%
  {\color{blue} #2}%
}
\newcommand\recalltext[1]{%
  \new{\ifcsname mytext@#1\endcsname
    \fontsize{10.5}{12.5}\selectfont\@nameuse{mytext@#1}%
  \else
    ``??''
  \fi
}}
\makeatother



\newacronym{3gpp}{3GPP}{3rd Generation Partnership Project}
\newacronym{adc}{ADC}{Analog to Digital Converter}
\newacronym{pdr}{PDR}{Packet Delivery Ratio}
\newacronym{4g}{4G}{4th generation}
\newacronym{5g}{5G}{5th generation}
\newacronym{6g}{6G}{6th generation}
\newacronym{ai}{AI}{Artificial Intelligence}
\newacronym{aimd}{AIMD}{Additive Increase Multiplicative Decrease}
\newacronym{am}{AM}{Acknowledged Mode}
\newacronym{tn}{TN}{Terrestrial Network}
\newacronym{amc}{AMC}{Adaptive Modulation and Coding}
\newacronym{aqm}{AQM}{Active Queue Management}
\newacronym{awgn}{AGWN}{Additive White Gaussian Noise}
\newacronym{balia}{BALIA}{Balanced Link Adaptation}
\newacronym{bdp}{BDP}{Bandwidth-Delay Product}
\newacronym{bf}{BF}{beamforming}
\newacronym{uu}{Uu}{Universal um}
\newacronym{refp}{RP}{Reference Point}
\newacronym{cc}{CC}{Congestion Control}
\newacronym{cdf}{CDF}{Cumulative Distribution Function}
\newacronym{cn}{CN}{Core Network}
\newacronym{cqi}{CQI}{Channel Quality Information}
\newacronym{ap}{AP}{Access Point}
\newacronym{cp}{CP}{Control Plane}
\newacronym{up}{UP}{User Plane}
\newacronym{upf}{UPF}{User Plane Function}
\newacronym{csirs}{CSI-RS}{Channel State Information - Reference Signal}
\newacronym{sib}{SIB}{System Information Block}
\newacronym{dc}{DC}{Dual Connectivity}
\newacronym{rb}{RB}{Resource Block}
\newacronym{dce}{DCE}{Direct Code Execution}
\newacronym{dci}{DCI}{downlink control onformation}
\newacronym{udp}{UDP}{User Datagram Protocol}
\newacronym{dl}{DL}{downlink}
\newacronym{drl}{DRL}{deep reinforcement learning}
\newacronym{fcfs}{FCFS}{first-come-first-served}
\newacronym{dmr}{DMR}{Deadline Miss Ratio}
\newacronym{fspl}{FSPL}{free-space path loss}
\newacronym{dmrs}{DMRS}{DeModulation Reference Signal}
\newacronym{e2e}{E2E}{End-to-End}
\newacronym{ppp}{PPP}{Poission Point Process}
\newacronym{aoi}{AoI}{Area of Interest}
\newacronym{cpu}{CPU}{Central Processing Unit}
\newacronym{gpu}{GPU}{Graphics Processing Unit}
\newacronym{tpu}{TPU}{Tensor Processing Unit}
\newacronym{ta}{TA}{Timing Advance}
\newacronym{si}{SI}{Study Item}
\newacronym{ecn}{ECN}{Explicit Congestion Notification}
\newacronym{edf}{EDF}{Earliest Deadline First}
\newacronym{enb}{eNB}{eNodeB}
\newacronym{epc}{EPC}{Evolved Packet Core}
\newacronym{es}{ES}{Edge Server}
\newacronym{cav}{CAV}{Connected and Autonomous Vehicle}
\newacronym{fdma}{FDMA}{Frequency Division Multiple Access}
\newacronym{fdd}{FDD}{Frequency Division Duplexing}
\newacronym{tdm}{TDM}{Time Division Multiplexing}
\newacronym{upa}{UPA}{Uniform Planar Array}
\newacronym{car}{CAR}{Circular Aperture Reflector }
\newacronym[firstplural=Radio Access Technologies (RATs)]{rat}{RAT}{Radio Access Technology}
\newacronym[firstplural=Radio Access Technology (RTs)]{rt}{RT}{Radio Technology}
\newacronym{fs}{FS}{Fast Switching}
\newacronym{isd}{ISD}{inter-site distance}
\newacronym{ftp}{FTP}{File Transfer Protocol}
\newacronym{gnb}{gNB}{Next Generation NodeB}
\newacronym{harq}{HARQ}{Hybrid Automatic Repeat reQuest}
\newacronym{hetnet}{HetNet}{Heterogeneous Network}
\newacronym{hh}{HH}{Hard Handover}
\newacronym{hol}{HOL}{Head-of-Line}
\newacronym{ia}{IA}{Initial Access}
\newacronym{imt}{IMT}{International Mobile Telecommunication}
\newacronym{iot}{IoT}{Internet of Things}
\newacronym{los}{LOS}{Line of Sight}
\newacronym{lte}{LTE}{Long Term Evolution}
\newacronym{m2m}{M2M}{Machine to Machine}
\newacronym{mac}{MAC}{Medium Access Control}
\newacronym{mc}{MC}{Multi-Connectivity}
\newacronym{mcs}{MCS}{Modulation and Coding Scheme}
\newacronym{mec}{MEC}{Mobile Edge Cloud}
\newacronym{mi}{MI}{Mutual Information}
\newacronym{mimo}{MIMO}{Multiple Input Multiple Output}
\newacronym{mmwave}{mmWave}{millimeter wave}
\newacronym{mptcp}{MP-TCP}{Multipath TCP}
\newacronym{mr}{MR}{Maximum Rate}
\newacronym{mss}{MSS}{Maximum Segment Size}
\newacronym{mtd}{MTD}{Machine-Type Device}
\newacronym{mtu}{MTU}{Maximum Transmission Unit}
\newacronym{nfv}{NFV}{Network Function Virtualization}
\newacronym{vnf}{VNF}{Virtualization Network Function}
\newacronym{gv}{GV}{ground vehicle}
\newacronym{vec}{VEC}{Vehicular Edge Computing}
\newacronym{dn}{DN}{Data Network}
\newacronym{sdn}{SDN}{Software Defined Networking}
\newacronym{nlos}{NLOS}{Non Line of Sight}
\newacronym{nlosb}{NLOSb}{Building Non Line of Sight}
\newacronym{nlosv}{NLOSv}{Vehicle Non Line of Sight}
\newacronym{nr}{NR}{New Radio}
\newacronym{ofdm}{OFDM}{Orthogonal Frequency Division Multiplexing}
\newacronym{pdcch}{PDCCH}{Physical Downlonk Control Channel}
\newacronym{sctp}{SCTP}{Stream Control Transport Protocol}
\newacronym{sdap}{SDAP}{Service Data Adaptation Protocol}
\newacronym{pdcp}{PDCP}{Packet Data Convergence Protocol}
\newacronym{pdsch}{PDSCH}{Physical Downlink Shared Channel}
\newacronym{pdu}{PDU}{Packet Data Unit}
\newacronym{pf}{PF}{Proportional Fair}
\newacronym{pgw}{PGW}{Packet Gateway}
\newacronym{sgw}{SGW}{Serving Gateway}
\newacronym{phy}{PHY}{Physical}
\newacronym{pbch}{PBCH}{Physical Broadcast Channel}
\newacronym[plural=\gls{mme}s,firstplural=Mobility Management Entities (MMEs)]{mme}{MME}{Mobility Management Entity}
\newacronym{prb}{PRB}{Physical Resource Block}
\newacronym{pss}{PSS}{Primary Synchronization Signal}
\newacronym{pucch}{PUCCH}{Physical Uplink Control Channel}
\newacronym{pusch}{PUSCH}{Physical Uplink Shared Channel}
\newacronym{rach}{RACH}{Random Access Channel}
\newacronym{ran}{RAN}{Radio Access Network}
\newacronym{red}{RED}{Random Early Detection}
\newacronym{rf}{RF}{Radio Frequency}
\newacronym{rlc}{RLC}{Radio Link Control}
\newacronym{rlf}{RLF}{Radio Link Failure}
\newacronym{rrc}{RRC}{Radio Resource Control}
\newacronym{rrm}{RRM}{Radio Resource Management}
\newacronym{rr}{RR}{Round Robin}
\newacronym{rs}{RS}{Remote Server}
\newacronym{rsrp}{RSRP}{Reference Signal Received Power}
\newacronym{rss}{RSS}{Received Signal Strength}
\newacronym{rtt}{RTT}{Round Trip Time}
\newacronym{rw}{RW}{Receive Window}
\newacronym{rx}{RX}{Receiver}
\newacronym{sa}{SA}{standalone}
\newacronym{sack}{SACK}{Selective Acknowledgment}
\newacronym{sap}{SAP}{Service Access Point}
\newacronym{sch}{SCH}{Secondary Cell Handover}
\newacronym{scoot}{SCOOT}{Split Cycle Offset Optimization Technique}
\newacronym{sdma}{SDMA}{Spatial Division Multiple Access}
\newacronym{sinr}{SINR}{Signal to Interference plus Noise Ratio}
\newacronym{sm}{SM}{Saturation Mode}
\newacronym{snr}{SNR}{Signal-to-Noise Ratio}
\newacronym{son}{SON}{Self-Organizing Network}
\newacronym{ss}{SS}{Synchronization Signal}
\newacronym{srs}{SRS}{Sounding Reference Signal}
\newacronym{sss}{SSS}{Secondary Synchronization Signal}
\newacronym{tb}{TB}{Transport Block}
\newacronym{tcp}{TCP}{Transmission Control Protocol}
\newacronym{tdd}{TDD}{Time Division Duplexing}
\newacronym{tdma}{TDMA}{Time Division Multiple Access}
\newacronym{tfl}{TfL}{Transport for London}
\newacronym{tm}{TM}{Transparent Mode}
\newacronym{prr}{PRR}{Packet Reception Ratio}
\newacronym{trp}{TRP}{Transmitter Receiver Pair}
\newacronym{tti}{TTI}{Transmission Time Interval}
\newacronym{ttt}{TTT}{Time-to-Trigger}
\newacronym{tx}{TX}{Transmitter}
\newacronym{ue}{UE}{User Equipment}
\newacronym{ul}{UL}{uplink}
\newacronym{uml}{UML}{Unified Modeling Language}
\newacronym{um}{UM}{Unacknowledged Mode}
\newacronym{utc}{UTC}{Urban Traffic Control}
\newacronym{vm}{VM}{Virtual Machine}
\newacronym{rsrq}{RSRQ}{Reference Signal Received Quality}
\newacronym{rssi}{RSSI}{Received Signal Strength Indicator}
\newacronym{crs}{CRS}{Cell Reference Signal}
\newacronym{v2v}{V2V}{Vehicle-to-Vehicle}
\newacronym{v2i}{V2I}{Vehicle-to-Infrastructure}
\newacronym{v2n}{V2N}{Vehicle-to-Network}
\newacronym{v2x}{V2X}{Vehicle-to-Everything}
\newacronym{vn}{VN}{Vehicular Node}
\newacronym{dsrc}{DSRC}{Dedicated Short Range Communication}
\newacronym{ci}{CI}{context information}
\newacronym{voi}{VoI}{value of information}
\newacronym{gps}{GPS}{Global Positioning System}
\newacronym{qos}{QoS}{Quality of Service}
\newacronym{qoe}{QoE}{Quality of Experience}
\newacronym{ml}{ML}{Machine Learning}
\newacronym{ahp}{AHP}{Analytic Hierarchy Process}
\newacronym{lidar}{LIDAR}{Light Detection and Ranging}
\newacronym{sumo}{SUMO}{Simulation of Urban MObility}
\newacronym{wave}{WAVE}{Wireless Access in Vehicular Environment}
\newacronym{c-its}{C-ITS}{Connected Intelligent Transportation System}
\newacronym{dash}{DASH}{Dynamic Adaptive Streaming over HTTP}
\newacronym{http}{HTTP}{HyperText Transfer Protocol}
\newacronym{nt}{NT}{Non-Terrestrial}
\newacronym{ntc}{NTC}{non-terrestrial communication}
\newacronym{ntn}{NTN}{Non-Terrestrial Network}
\newacronym{haps}{HAPS}{High Altitude Platform Station}
\newacronym{hap}{HAP}{High Altitude Platform}
\newacronym{leo}{LEO}{Low Earth Orbit}
\newacronym{meo}{MEO}{Medium Earth Orbit}
\newacronym{geo}{GEO}{Geostationary Earth Orbit}
\newacronym{uav}{UAV}{Unmanned Aerial Vehicle}
\newacronym{nsat}{nSAT}{Nanosatellite}
\newacronym{ehf}{EHF}{extremely high-frequency}
\newacronym{ioe}{IoE}{Internet of Everyone}
\newacronym{gan}{GaN}{Gallium Nitride}
\newacronym{af}{AF}{amplify-and-forward}
\newacronym{csi}{CSI}{channel state information}
\newacronym{ecdf}{ECDF}{empirical cumulative distribution function}
\newacronym{f}{F}{flexible}
\newacronym{fpga}{FPGA}{field programmable gate array}
\newacronym{fov}{FoV}{field-of-view}
\newacronym{km}{KM}{K-means}
\newacronym{kmed}{KMed}{K-medoids}
\newacronym{iab}{IAB}{Integrated Access and Backhaul}
\newacronym{bap}{BAP}{backhaul adaptation protocol}
\newacronym{irs}{IRS}{intelligent reflecting surface}
\newacronym{lsfc}{LSFC}{large-scale fading coefficient}
\newacronym{noma}{NOMA}{non-orthogonal multiple access}
\newacronym{fdm}{FDM}{frequency-division multiplexing}
\newacronym{sdm}{SDM}{space-division multiplexing}
\newacronym{ofdma}{OFDMA}{orthogonal frequency-division multiple access}
\newacronym{oma}{OMA}{orthogonal multiple access}
\newacronym{plos}{pLoS}{probabilistic \ac{los}}
\newacronym{rsma}{RSMA}{rate-splitting multiple access}
\newacronym{scm}{SCM}{spatial channel model}
\newacronym{siso}{SISO}{single input single output}
\newacronym{svd}{SVD}{singular value decomposition}
\newacronym{thz}{THz}{Terahertz}
\newacronym{ula}{ULA}{uniform linear array}
\newacronym{uma}{UMa}{urban macro-cell}
\newacronym{umi}{UMi}{urban micro-cell}
\newacronym{mt}{MT}{mobile terminal}
\newacronym{cu}{CU}{centralized unit}
\newacronym{du}{DU}{distributed unit}
\newacronym{dag}{DAG}{directed acyclic graph}
\newacronym{st}{ST}{spanning tree}
\newacronym{rma}{RMa}{rural macrocell}
\newacronym{inf}{InF}{indoor factory}
\newacronym{ngc}{NGC}{next generation core}
\newacronym{gtp}{GTP}{GPRS Tunnelling Protocol}
\newacronym{tft}{TFT}{Traffic Flow Template}
\newacronym{teid}{TEID}{Tunnel Endpoint Identifier}
\newacronym{tnl}{TNL}{Transport Network Layer}
\newacronym{amf}{AMF}{Access and Mobility Management Function}
\newacronym{ngso}{NGSO}{Non-Geostationary Orbit}
\newacronym{redcap}{RedCap}{Reduced Capability}
\newacronym{ng}{NG}{Next Generation}
\newacronym{fr1}{FR1}{Frequency Range 1}
\newacronym{prach}{PRACH}{Physical Random Access Channel}
\newacronym{ro}{RO}{RACH Occasion}
\newacronym{app}{APP}{Application}
\newacronym{vsat}{VSAT}{Very Small Aperture Terminal}
\newacronym{ack}{ACK}{Acknowledgment}
\newacronym{cwnd}{CWND}{Congestion Window}
\newacronym{rar}{RAR}{Random Access Response}
\newacronym{n6}{N6}{Network 6}
\newacronym{rto}{RTO}{Retransmission Timeout}
\newacronym{ims}{IMS}{IP Multimedia Subsystem}
\newacronym{6gr}{6GR}{6G Radio}
\newacronym{sar}{SAR}{search-and-rescue}
\newacronym{ps}{PS}{Public Safety}
\newacronym{ppdr}{PPDR}{Public Protection and Disaster Relief}

\title{Experimental Evaluation of a UAV-Mounted LEO Satellite Backhaul for Emergency Connectivity}

\author{\IEEEauthorblockN{Mattia Figaro$^\star$, Francesco Rossato$^\star$, Alexander Bonora$^\star$, Marco~Giordani$^\star$, Giovanni Schembra$^\dagger$, Michele Zorzi$^\star$\\ }
\IEEEauthorblockA{$^\star$Department of Information Engineering (DEI), University of Padova, Italy.\\
Emails: \{mattia.figaro, francesco.rossato, marco.giordani, alexander.bonora, michele.zorzi\}@dei.unipd.it.}\\
$^\dagger$ University of Catania. Email: giovanni.schembra@unict.it.}
\begin{document}
\maketitle

\begin{abstract}
Reliable connectivity is critical for \gls{ppdr} operations, especially in rural or compromised environments where terrestrial infrastructure is unavailable. 
In such scenarios, \glspl{ntn}, and specifically \glspl{uav}, are promising candidates to provide on-demand and rapid connectivity on the ground, serving as aerial base stations.
In this paper, we implement a setup in which a rotary-wing \gls{uav}, equipped with a Starlink Mini terminal, provides Internet connectivity to an emergency ground user in the absence of cellular coverage via \gls{leo} satellites. The UAV functions as a Wi-Fi access point, while backhauling the ground traffic through the Starlink constellation.
We evaluate the system via both network simulations in ns-3 and real-world flight experiments in a rural environment, in terms of throughput, latency, coverage, and energy consumption under static and dynamic flight conditions. 
Our results demonstrate that the system can maintain a stable uplink throughput of approximately 30 Mbps up to approximately 200 meters, and with minimal impact on the UAV battery lifetime. 
These findings demonstrate the feasibility of deploying commercial LEO satellite terminals on UAVs as a practical solution for emergency connectivity.
\end{abstract}
\glsresetall

\begin{IEEEkeywords}
    \glspl{uav}; \glspl{ntn}; \gls{leo} satellites; Starlink; emergency communications; experimental analysis.
\end{IEEEkeywords}

\begin{tikzpicture}[remember picture,overlay]
\node[anchor=north,yshift=-10pt] at (current page.north) {\parbox{\dimexpr\textwidth-\fboxsep-\fboxrule\relax}{
\centering\footnotesize This paper has been accepted for presentation at the 2026 International Conference on Computing, Networking and Communications (ICNC). \textcopyright 2026 IEEE. \\
Please cite it as: M. Figaro, F. Rossato, A. Bonora, M. Giordani, G. Schembra M. Zorzi, ``Experimental Evaluation of a UAV-Mounted LEO Satellite Backhaul for Emergency Connectivity'' in International Conference on Computing, Networking and Communications (ICNC), 2026.\\
}};
\end{tikzpicture}

\glsresetall
\section{Introduction}
\label{sec:introduction}
Communication networks play a crucial role in \gls{ppdr} operations, to protect lives and ensure public safety, enable first responders and authorities to rapidly coordinate emergency interventions, and allocate personnel and equipment efficiently.
As the number and severity of natural disasters (e.g., fires, floods,
earthquakes, tsunamis, volcano eruptions) increases globally, and as armed conflicts and humanitarian crises evolve in scale and complexity, robust~\gls{ppdr} systems are vital to maintain reliable communication during emergencies~\cite{karaman2025solutions,di2024emerging}.

In recent years, \glspl{ntn}~\cite{giordani2020non} based on aerial and spaceborne platforms have emerged as a key component of \gls{6g} wireless systems~\cite{giordani2020toward} to provide wide-area coverage beyond the limits of terrestrial cellular networks, e.g., in rural and/or remote areas, and rapidly restore connectivity in disaster areas where ground infrastructure is either absent or compromised~\cite{chaoub20216g}. 
In particular, \glspl{uav} such as drones can act as flexible, fast-to-deploy, plug-and-play aerial base stations and edge nodes~\cite{traspadini2022uav}, to dynamically provide access connectivity to ground nodes in hard-to-reach areas where first responders cannot easily operate~\cite{kuccukerdem2025autonomous,boschiero2020coverage}. 
Another approach for \gls{ppdr} is to rely on \gls{leo} satellites.
Unlike \gls{geo} satellites, \gls{leo} satellites operate much closer to the Earth, typically between 300 and 1\,000 km, which enables similar latency and throughput as in terrestrial networks~\cite{giordani2020satellite}.
Notably, LEO satellites can provide direct Internet connectivity by either relaying data to ground gateways (transparent payload) or processing data onboard (regenerative payload), making them particularly attractive for emergency connectivity~\cite{figaro20255g}.
In this context, Starlink, serving more than 8 million customers as of November 2025, represents one of the most mature examples of satellite-based Internet solutions, providing service performance comparable to that of many 4G and 5G cellular operators, particularly in rural or underserved areas~\cite{lagunas2024low,michel2022first}. 

The scientific community is also exploring the integration of \gls{leo} satellite terminals directly onto rotary-wing \glspl{uav}. 
Specifically, \glspl{uav} can serve as aerial access points for ground users, and ultimately relay data traffic through the LEO satellite network to the Internet. 
For instance, Almeida \emph{et al.}~\cite{almeida2024uav} demonstrated a framework where a \gls{uav} equipped with object detection capabilities served as an aerial mobile 5G base station via Starlink. While the system was able to support direct-to-smartphone connectivity, the satellite antenna was not mounted on the drone itself but rather on  the ground, so the UAV could only operate within the immediate vicinity of the ground station.
Conversely, Jordan \emph{et al.} \cite{jordan2025successful} successfully deployed a Starlink system on an Ultra \gls{uav} (a large fixed-wing platform) for scientific data collection in Antarctica. Despite the mission’s success, the authors noted that significant design iterations were required to reduce weight and power consumption for routine operations. Similarly, Liu \emph{et al.} \cite{liu2025starlink} tested Starlink onto a Unitree GO2 Robot Dog, providing valuable datasets for satellite communication, even though the dynamics, energy, and payload constraints of a wheeled robot are fundamentally different from those of a rotary-wing UAV.

Compared to prior work, in this paper we implement a setup in which a Starlink Mini is directly mounted onto a commercial rotary-wing X8-1000 Pro RTK UAV. Specifically, a ground user connects to the UAV via Wi-Fi, while Internet access is provided through the Starlink constellation.
We test this system via real-world experiments in a rural area near Catania with no cellular coverage, measuring throughput, latency, coverage, and
energy consumption under static and dynamic flight conditions.
Our results demonstrate that the drone can support the weight of the Starlink antenna, reducing the flight time by only about 8\%.
Moreover, we measure a stable uplink throughput of approximately 30 Mbps over a range of 200 meters, thereby demonstrating that this setup provides a stable and robust backhaul connectivity solution for \gls{ppdr}.
Our experimental results have been validated by laboratory simulations using \texttt{ns3-NTN}, an open-source ns-3 module that simulates full-stack satellite communication based on the \gls{3gpp} specifications~\cite{sandri23implementation}. 

The paper is organized as follows. Sec.~\ref{sec:scenario_and_objectives} describes our scenario and motivations, Sec.~\ref{sec:experimental_setup} presents the experimental and simulation setups, Sec.~\ref{sec:results} presents our test results, and Sec.~\ref{sec:conclusions} concludes the paper with suggestions for future work.


\begin{figure*}[t!]
    \centering
    \includegraphics[width=0.95\textwidth]{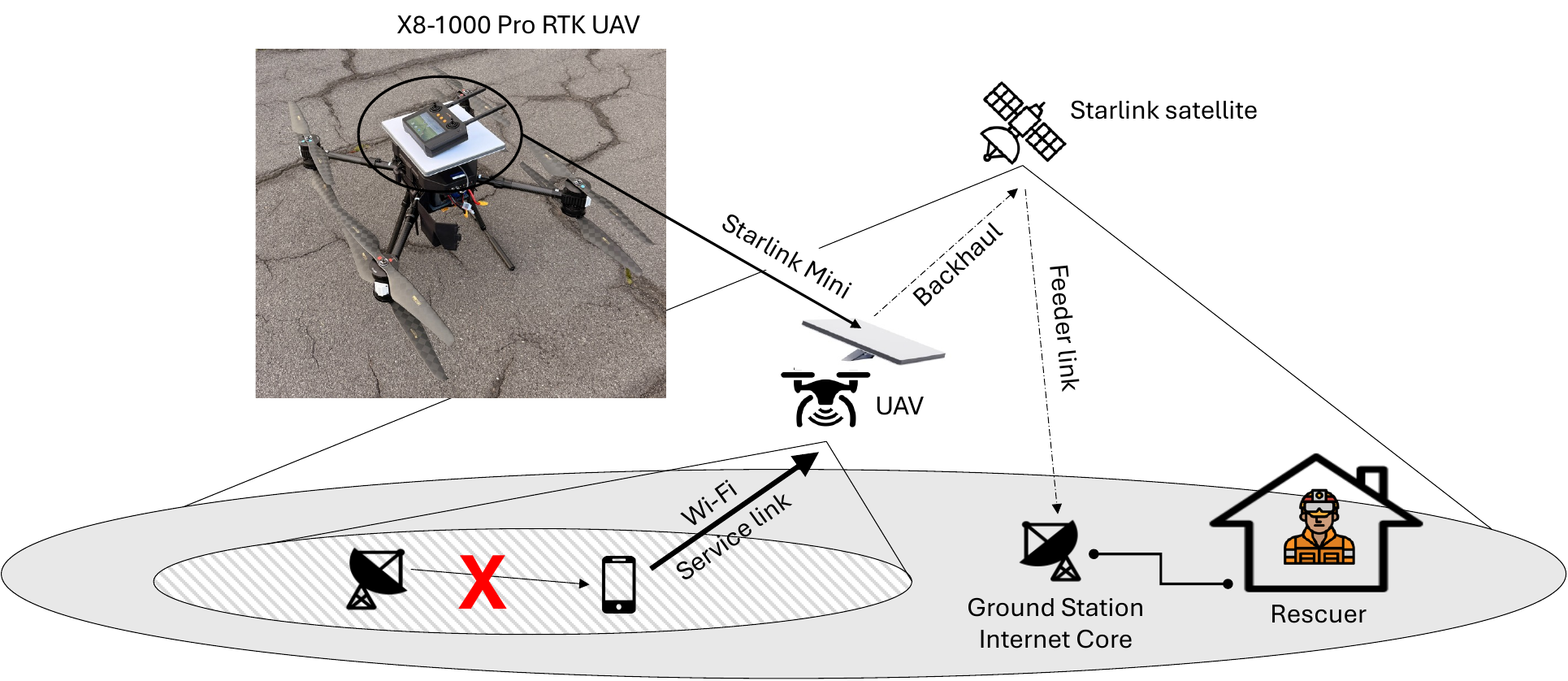}
    \caption{A disaster-relief \gls{ppdr} scenario in which a rotary-wing UAV drone, mounting a Starlink Mini antenna terminal, relays data traffic from a ground user via Wi-Fi to the rescuer through a Starlink satellite backhaul link.}
    \label{fig:disaster_relief}
\end{figure*}

\section{Scenario and Objectives}
\label{sec:scenario_and_objectives}

\subsection{Motivations}
In a \gls{ppdr} scenario, sudden natural or human-made events compromise traditional communication infrastructures, particularly power lines, fiber backhaul links, base stations, and core network nodes. Such disruptions are particularly severe in rural or otherwise underserved regions where network redundancy is limited. 
For example, the 6.2-magnitude earthquake that struck central Italy in August 2016 devastated towns such as Amatrice and Norcia, leaving them without telephone or Internet access for several days due to damaged power lines and telecommunication cables~\cite{shaffiee2025potential}.
More recently, in May 2023, severe flooding and landslides in the Emilia-Romagna region of Italy resulted in Internet outages for more than 30,000 users in the affected areas~\cite{bakhtyari2025impact}, and impaired rescue~operations.

In this scenario, \glspl{uav} may serve as mobile base stations from the sky~\cite{bordin2022autonomous}, providing a reliable local network for \gls{ppdr} operators and ground teams when terrestrial infrastructure is unavailable.
Moreover, \glspl{uav} can carry sensing payloads such as optical, infrared, or thermal cameras to assist \gls{sar} operations by detecting survivors or other points of interest in real time \cite{uav_sar}.
On top of this, \glspl{uav} can provide on-demand connectivity for other time-critical or infrastructure-limited applications, such as remote industrial inspections, environmental monitoring, and agricultural or forestry management, while collecting and transmitting relevant data in real time. 
However, UAVs generally require a reliable \gls{los} connection link with a nearby terrestrial base station to relay the access ground traffic or local sensing data to the Internet, which is often unavailable in disaster scenarios~\cite{shi2018drone}.

Satellites can also provide robust, wide-area, and uninterrupted Internet connectivity during emergencies. For example, during recent weather disasters in Italy, Starlink satellites were reported to support an average Internet throughput of around 50 Mbps~\cite{shaffiee2025potential}, up to three times faster than ground networks in disrupted zones. 
However, current architectures require user terminals to mount relatively large antenna panels (e.g., Starlink terminals) to connect to the satellites, which cannot be integrated into commercial user devices such as smartphones, thereby making direct-to-device standalone satellite connectivity difficult to achieve.

\subsection{Our \gls{ppdr} Scenario}
To solve the issues of standalone UAV or satellite communication systems, we consider the \gls{ppdr} scenario illustrated in Fig.~\ref{fig:disaster_relief}.
Specifically, we consider a multi-role rotary-wing UAV drone with both sensing and communication capabilities. In its communication role, it acts as a mobile aerial Wi-Fi gateway providing access connectivity to ground users located in disaster areas where terrestrial networks are unavailable.
To establish a backhaul link to the Internet, otherwise unavailable, the UAV carries a Starlink Mini antenna terminal in the payload, which relays the ground traffic to a Starlink satellite. From there, data is transmitted to a ground station and ultimately routed to the Internet, to support \gls{ppdr} operations in the rescue center.\footnote{It is not publicly known whether Starlink satellites operate in a bent-pipe mode (i.e., connecting directly to ground gateways via transparent payloads) as in first-generation satellites, or they leverage regenerative payload techniques (which would make it possible to use inter-satellite links for traffic routing) as seen in second-generation satellites.  For simplicity, in  Fig.~\ref{fig:disaster_relief} we depict the former option.} 

The primary goal of our research is to evaluate the operational limits and feasibility of an integrated \gls{uav}-satellite mobile gateway system as a practical solution to provide emergency Internet connectivity. We will demonstrate via field experiments and computer simulations that transporting a Starlink Mini antenna on a drone is possible in terms of payload weight and capacity, energy consumption, aerodynamics, and communication performance.


\begin{figure*}
    \centering
    \includegraphics[width=0.99\textwidth]{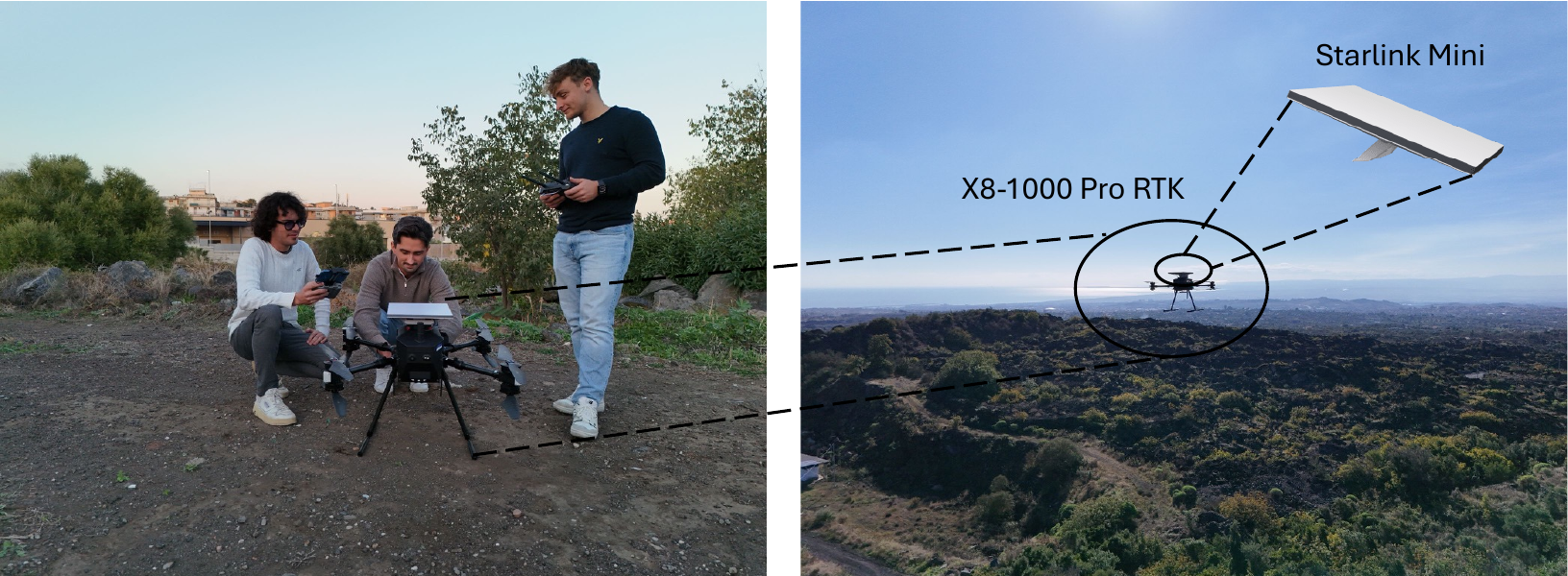}
    \caption{Some of the authors of the paper with the X8-1000 Pro RTK UAV drone, mounting a Starlink Mini antenna terminal (left). An aerial overview of the experimental rural area near Catania (right).}
    \label{fig:fig2}
\end{figure*}

\section{Experimental and Simulation Setup}
\label{sec:experimental_setup}

In this section we describe our experimental (Sec.~\ref{sub:ex}) and simulation setup (Sec.~\ref{sub:sim}). 

\subsection{Experimental Setup}
\label{sub:ex}

Our experimental setup, illustrated in Fig.~\ref{fig:fig2}, is conceptually designed as a rapidly deployable flying backhaul system for disaster areas where traditional terrestrial networks are damaged or offline.
The architecture integrates a commercial off-the-shelf \gls{uav} with a \gls{leo} satellite antenna terminal to create a mobile relay node. 

\paragraph{UAV}
The aerial node is a DroneBase X8-1000 Pro RTK octocopter.\footnote{DroneBase Professional: ~\url{https://dronebase.it/en/professional-drones/}.} This platform was selected for its high payload capacity (up to 10~kg) and strong stability in adverse conditions, including resistance to high winds (up to 35~knots), which are essential requirements for maintaining reliable satellite alignment.
Moreover, the drone utilizes a Dual-Antenna RTK GNSS module to achieve centimeter-level positioning accuracy, a critical capability for \gls{ppdr} applications such as photogrammetry, mapping, and \gls{sar} missions.
To ensure sufficient flight endurance, the propulsion system is powered by two high-voltage lithium-polymer (LiPoHV) batteries with a 6S configuration (22.8 V) and a capacity of 30\,000 mAh each. This configuration enables competitive flight times of up to 50 minutes, even when operating at full payload~capacity.

\paragraph{LEO satellite antenna terminal}
The communication payload consists of a flat-panel Starlink Mini antenna.\footnote{Starlink Mini: \url{https://starlink.com/specifications}.} This terminal was selected for its compact dimensions ($298.5\times259\times38.5$~mm), low weight (1.16~kg), and reduced power consumption (25-40~W), so it can be integrated within the drone's maximum takeoff weight constraints. 
The antenna is an Electronic Phased Array (EPA), which eliminates the need for mechanical steering, and permits to track multiple Starlink satellites simultaneously within a 110° field of view.
Starlink Mini performs a dual function: it provides a backhaul link to the Starlink \gls{leo} satellite constellation, and simultaneously acts as a Wi-Fi access point broadcasting a local network for ground users. The Wi-Fi module is a Dual Band Wi-Fi 5 (802.11a/b/g/n/ac) router, supporting coverage to up to 128 connected devices, across an area of up to 112~m$^2$.

At its core, the Starlink constellation, operated by SpaceX, currently consists of around 9\,000 \gls{leo} satellites, deployed across multiple orbital shells with different inclinations, at altitudes around 550-570~km. It enables near-global coverage, with an average downlink (uplink) speed of around 110 (15) Mbps and \glspl{rtt} below 40~ms~\cite{laniewski2025measuring}. 
The satellites operate primarily in the Ku-band at 10.7–12.7 GHz (downlink) and 14–14.5 GHz (uplink), and in the Ka-band at 17.8–18.6 GHz and 18.8–19.3 GHz (downlink) and 27.5–30 GHz (uplink), subject to regional authorisations~\cite{mohan2024multifaceted}.

\paragraph{Ground \gls{ue}}
The user segment consists of a Lenovo Yoga S730 laptop, which connects to the aerial node via standard Wi-Fi protocols. 

\subsection{Simulation Setup}
\label{sub:sim}

To validate on-the-field flight measurements, we reproduced our target scenario in ns-3, a widely adopted and reliable framework for end-to-end network simulation. Specifically, we use the \texttt{ns3-NTN} module~\cite{sandri23implementation},\footnote{The source code of the module: \url{https://gitlab.com/mattiasandri/ns-3-ntn}.} an open-source extension of ns-3 that we developed to simulate full-stack satellite communication according to the \gls{3gpp} 5G NR-\gls{ntn} Release 17$+$ specifications. This module implements the \gls{3gpp} \gls{ntn} channel model based on \cite{38811}, incorporating path loss, atmospheric absorption, scintillation, and frequency-dependent fading in the S-, L-, and Ka-bands. Moreover it includes antenna models for circular aperture, \gls{vsat}, and \gls{upa} configurations from \cite{38821}, and an \gls{ntn}-oriented Earth-Centered, Earth-Fixed (ECEF) Cartesian coordinate system. 
It also implements timing advance mechanisms~\cite{38214} and customized \gls{rrc} and \gls{harq} timers~\cite{38821} for precise propagation delay computation. The accuracy of \texttt{ns3-NTN} has been validated against 3GPP calibration results in~\cite{sandri23implementation,figaro20255g}, and currently appears as one of the most reliable and accessible software tools for NTN simulations.

In this work, the module has been further extended to simulate the multi-layer communication setup considered in this paper, specifically for Earth-UAV-satellite links.

\section{Real-World Tests and Results}
\label{sec:results}

Experimental tests were conducted in a rural area near Catania (Sicily, Italy) with minimal radio frequency interference and quasi-perfect \gls{los} between the \gls{uav} and the ground \gls{ue}, as illustrated in Fig.~\ref{fig:fig2} (right).
All flight tests were performed under optimal weather conditions, characterized by clear skies and minimal cloud cover, to establish a performance baseline with negligible atmospheric attenuation or rain fade. 

We evaluate the system's performance in terms of throughput, \gls{rtt}, signal stability, and energy consumption, using the iperf3 network testing tool.
Field measurements were collected under both static and dynamic flight conditions, where the UAV was either in stationary hovering or in constant motion, at a constant height of 20 meters.



\paragraph{Static hovering}

In Fig.~\ref{fig:thr_barplot} we evaluate the average \gls{dl} and \gls{ul} throughput when the \gls{uav} is in stationary hovering, gradually increasing the horizontal distance between the ground UE and the UAV up to 250 m. In these tests, we quantify the link degradation over distance to determine the maximum effective range of the system. 

First, we observe that there is a close match between the experimental (solid bars) and simulation results with \texttt{ns3-NTN} (striped bars), which demonstrates the mutual accuracy of our simulator and experimental platform.
For the \gls{dl} case, the maximum throughput is around 90 Mbps within the 0-50 m range, corresponding to optimal link conditions, which is consistent with typical Starlink network performance reported in previous studies~\cite{laniewski2025measuring}. 
Then, the throughput starts to decrease as the horizontal distance between the ground UE and the UAV increases, dropping to around 30 Mbps in the 150-200 m range, and the link is completely lost beyond 200 m. 
This behavior highlights that, when the \gls{ue} and the \gls{uav} are too far (more than 200 m in our results), the performance degrades both in UL and DL due to the failure of the Wi-Fi link, as expected. On the other hand, for shorter distances the Wi-Fi link works well, but a different behavior is observed in the two directions. Specifically, in the DL the throughput keeps increasing as the distance decreases, whereas in the UL it saturates at around 25-30 Mbps, showing that in this case the bottleneck is due to the limited Starlink capacity. Despite this difference, these results demonstrate that it is possible to establish effective DL and UL Internet connectivity to and from the ground via satellites, using a \gls{uav} both as a relay and as a backhaul node.

\begin{figure}[t!]
    \centering
    \definecolor{dl}{RGB}{112,105,157}
\definecolor{ul}{RGB}{197,105,161}

\definecolor{dl_simulated}{RGB}{159, 152, 196}
\definecolor{ul_simulated}{RGB}{232, 162, 205}

\usetikzlibrary{positioning}

\begin{tikzpicture}

\begin{axis}[
    width=8cm,
    name=data,
    ybar=0pt, 
    bar width=12pt,
    enlarge x limits=0.1, 
    ylabel={Throughput [Mbps]},
    xlabel={Horizontal UE-UAV distance [m]},
    xtick={1, 2, 3, 4, 5},
    xticklabels={0-50, 50-100, 100-150, 150-200, 200-250},
    ymin=0, ymax=100,
    ymajorgrids=true,
    grid style=dashed,
    tick label style={font=\footnotesize},
    ylabel style={font=\footnotesize},
    xlabel style={font=\footnotesize}
]

\addplot[fill=dl, draw=black] coordinates {(1, 92) (2, 57) (3, 40.5) (4, 28) (5, 0.2)};
\addplot[fill=ul, draw=black] coordinates {(1, 29) (2, 19.5) (3, 23.5) (4, 17) (5, 1)};
\end{axis}

\begin{axis}[
    at={(data.south west)},
    anchor=south west,
    width=8cm,
    ybar=0pt, 
    bar width=8pt,
    enlarge x limits=0.1, 
    ylabel={Throughput [Mbps]},
    xtick={1, 2, 3},
    xticklabels={LEO 600 km, LEO 1200 km, GEO 36000 km},
    ymin=0, ymax=100,
    ymajorgrids=true,
    grid style=dashed,
    tick label style={font=\footnotesize},
    ylabel style={font=\footnotesize},
    xlabel style={font=\footnotesize},
    hide axis]
\addplot[fill=dl_simulated, bar width=5pt, bar shift=-6pt, draw=black, postaction={pattern=north east lines}] coordinates
    {(1, 95.5) (2, 75) (3, 43.09) (4, 23.21) (5, 0)};
\addplot[fill=ul_simulated, bar width=5pt, bar shift=6pt, draw=black, postaction={pattern=north east lines}] coordinates
    {(1, 28.7) (2, 28.7) (3, 28.7) (4, 21.6) (5, 0)};
\end{axis}

\node[
    above=5pt of data.north, 
    anchor=south, 
    draw, 
    fill=white, 
    inner sep=3pt, 
    font=\small, 
] (customlegend) {
    {
    \setlength{\tabcolsep}{2pt}
    \begin{tabular}{c c c c}
    \tikz \draw[fill=dl, draw=black] (0,0) rectangle (0.3,0.3); & Experiments (DL) &
    \tikz \draw[fill=ul, draw=black] (0,0) rectangle (0.3,0.3); & Experiments (UL) \\
    \tikz \draw[draw=black, fill=dl_simulated, postaction={pattern=north east lines}] (0,0) rectangle (0.3,0.3); & Simulations (DL) &
    \tikz \draw[draw=black, fill=ul_simulated, postaction={pattern=north east lines}] (0,0) rectangle (0.3,0.3); & Simulations (UL)
    \end{tabular}
    }
};

\end{tikzpicture}
    \caption{Average DL and UL throughput in static hovering vs. the horizontal distance between the \gls{ue} and the \gls{uav}. Solid bars are for real-world experiments, while striped bars are for ns-3 simulations.}
    \label{fig:thr_barplot}
\end{figure}
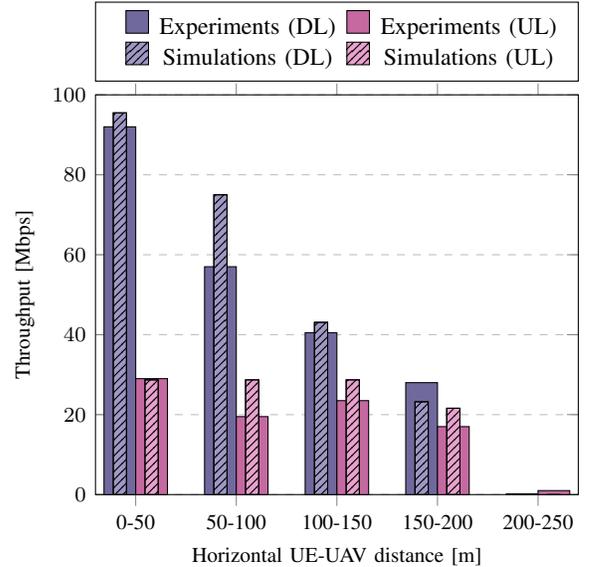 

\begin{figure}[t!]
    \centering
\begin{tikzpicture}
\tikzset{every node/.style={font=\footnotesize}}
\definecolor{dl}{RGB}{112,105,157}
\definecolor{ul}{RGB}{197,105,161}

\begin{axis}[
tick align=outside,
tick pos=left,
xlabel={Time [s]},
xmajorgrids,
    xtick={0, 10, ..., 120},
    ylabel={Throughput [Mbps]},
    ymajorgrids,
    legend pos=north east,
    legend columns=2,   
    ymin=0, ymax=20,
    ymajorgrids=true,
    grid style=dashed,
    tick label style={font=\footnotesize},
    ylabel style={font=\footnotesize},
    xlabel style={font=\footnotesize},
]

\addplot [line width=1pt, dl,  mark=x, mark size=2, mark options={fill=none}]
coordinates {%
(1.000542, 0)
(2.000564, 0)
(3.000447, 0)
(4.000495, 0)
(5.000169, 0)
(6.000558, 0)
(7.000476, 0)
(8.000513, 0)
(9.000594, 0)
(10.000561, 0)
(11.000577, 0)
(12.000542, 0)
(13.000592, 0)
(14.000466, 0)
(15.000517, 0)
(16.000139, 0)
(17.000528, 0)
(18.000543, 0)
(19.000215, 0)
(20.000556, 0.104888)
(21.000513, 0.104888)
(22.000564, 0.209751)
(23.000528, 0.314603)
(24.000545, 0.52432)
(25.000298, 0.734035)
(26.000517, 0.943749)
(27.000527, 1.363323)
(28.000569, 1.468141)
(29.000528, 1.677864)
(30.000297, 1.677833)
(31.000577, 2.306976)
(32.000529, 3.146051)
(33.000509, 3.250868)
(34.000519, 3.985082)
(35.000364, 4.509195)
(36.000249, 5.662717)
(37.000147, 5.452846)
(38.000148, 6.396604)
(39.000164, 7.654762)
(40.000558, 9.12303)
(41.000149, 9.856902)
(42.00033, 8.912964)
(43.000509, 9.332485)
(44.000435, 8.388553)
(45.000572, 8.388499)
(46.000444, 7.339955)
(47.000533, 7.549533)
(48.000576, 7.549564)
(49.000596, 7.130237)
(50.000592, 7.025519)
(51.000555, 6.291501)
(52.000514, 6.501204)
(53.000221, 5.872015)
(54.000516, 5.872015)
(55.000467, 6.711101)
(56.000522, 8.808179)
(57.000513, 11.744274)
(58.000513, 12.373505)
(59.00052, 11.324951)
(60.000177, 10.905191)
(61.000553, 11.220142)
(62.000519, 15.098597)
(63.000516, 16.252343)
(64.000521, 16.566822)
(65.000288, 14.993963)
(66.000513, 13.525933)
(67.000287, 13.422167)
(68.000526, 12.792645)
(69.000462, 13.841447)
(70.00052, 14.99417)
(71.00052, 14.050223)
(72.000606, 11.220039)
(73.000439, 10.590574)
(74.000538, 11.219756)
(75.000585, 11.849086)
(76.000515, 10.905345)
(77.000222, 7.863437)
(78.000535, 8.59827)
(79.000155, 10.064988)
(80.000512, 9.961237)
(81.000483, 10.380646)
(82.000504, 9.961338)
(83.00023, 10.402529)
(84.000468, 9.458867)
(85.000526, 8.829538)
(86.000527, 8.829538)
(87.000529, 8.619836)
};
\addlegendentry{DL}

\addplot [line width=1pt, ul,  mark=x, mark size=2, mark options={fill=none}]
coordinates{
(1, 0)
(2.52631578947368, 0)
(4.05263157894737, 0)
(5.57894736842105, 0)
(7.10526315789474, 0)
(8.63157894736842, 0)
(10.1578947368421, 0)
(11.6842105263158, 0.427514)
(13.2105263157895, 0.427514)
(14.7368421052632, 0.427514)
(16.2631578947368, 0.427514)
(17.7894736842105, 0.427514)
(19.3157894736842, 0.427514)
(20.8421052631579, 0.427514)
(22.3684210526316, 0.427514)
(23.8947368421053, 0.427514)
(25.421052631579, 0.427514)
(26.9473684210526, 0.325344)
(28.4736842105263, 0.325344)
(30, 0.325344)
(31.5263157894737, 0.325344)
(33.0526315789474, 0.325344)
(34.5789473684211, 0.954233)
(36.1052631578947, 1.583304)
(37.6315789473684, 2.212628)
(39.1578947368421, 2.946312)
(40.6842105263158, 3.575776)
(42.2105263157895, 3.984153)
(43.7368421052632, 4.823339)
(45.2631578947368, 5.452685)
(46.7894736842105, 6.081533)
(48.3157894736842, 6.815854)
(49.8421052631579, 7.13069)
(51.3684210526316, 7.969647)
(52.8947368421053, 8.283764)
(54.421052631579, 7.65493)
(55.9473684210526, 7.025466)
(57.4736842105263, 6.291744)
(59, 5.66223)
(60.5263157894737, 5.557248)
(62.0526315789474, 5.557451)
(63.5789473684211, 6.260976)
(65.1052631578947, 9.930156)
(66.6315789473684, 8.671832)
(68.1578947368421, 7.833285)
(69.6842105263158, 7.728436)
(71.2105263157895, 7.728436)
(72.7368421052632, 7.728436)
(74.2631578947369, 7.518764)
(75.7894736842105, 6.9944)
(77.3157894736842, 6.470207)
(78.8421052631579, 5.241933)
(80.3684210526316, 2.075744)
(81.8947368421053, 3.312755)
(83.421052631579, 3.207861)
(84.9473684210526, 3.207861)
(86.4736842105263, 3.207861)
(88, 3.207861)
};
\addlegendentry{UL}

\end{axis}

\end{tikzpicture}
    
    \caption{Evolution of the DL and UL throughput during dynamic flight, as the \gls{uav} moves from an initial horizontal distance of 250~m toward the \gls{ue} at a constant speed of approximately 2.5 m/s.}
    \label{fig:thr_towards}
\end{figure}
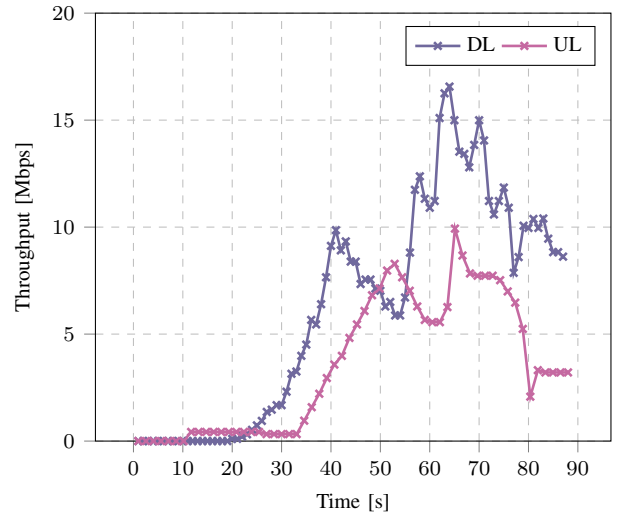

\paragraph{Dynamic mobility}
In these tests we assess the resilience of the system during active flight, with the \gls{uav} in constant motion. 
In Fig. \ref{fig:thr_towards} we plot the DL and UL throughput as the \gls{uav} moves from a horizontal distance of 250 m toward the \gls{ue} at constant velocity. We recognize different regimes. 
At first, the throughput is nearly zero since the Wi-Fi link is unavailable beyond 200 m, confirming our results in Fig.~\ref{fig:thr_barplot}.
Then, as the \gls{uav} approaches the \gls{ue}, the Wi-Fi link finally activates, and the DL (UL) throughput gradually increases up to around 18 (10) Mbps after around 65 s (when the distance to the drone is around 100 m).
Nevertheless, we observe that the throughput is quite unstable.
This behavior was also recorded in offline tests where the Starlink Mini antenna was mounted on the ground and connected to the \gls{ue} via a wired Ethernet link. Therefore, we conclude that this instability is inherent to the satellite backhaul itself, and independent of the UAV mobility or the local Wi-Fi~connection.

\begin{figure}[t!]
    \centering
\begin{tikzpicture}
\tikzset{every node/.style={font=\footnotesize}}

\definecolor{dl}{RGB}{112,105,157}
\definecolor{ul}{RGB}{197,105,161}


\definecolor{dl_simulated}{RGB}{159, 152, 196}
\definecolor{ul_simulated}{RGB}{232, 162, 205}

\begin{axis}[
tick align=outside,
tick pos=left,
xlabel={Time [s]},
xmajorgrids,
    xtick={0, 20, ..., 120},
    ylabel={Throughput [Mbps]},
    ymajorgrids,
    legend pos=north east,
    ymin=0, ymax=110,
    ymajorgrids=true,
    grid style=dashed,
    legend columns=3,
    legend style={at={(0.95,1.13)}},
    tick label style={font=\footnotesize},
    ylabel style={font=\footnotesize},
    xlabel style={font=\footnotesize}
]

\addplot [line width=1pt, dl, mark=x, mark size=2, mark options={fill=none}]
coordinates {
(1.001238, 17.928457)
(2.001243, 21.387606)
(3.001059, 24.744375)
(4.001123, 28.518958)
(5.001072, 32.397638)
(6.001405, 36.381797)
(7.001010, 37.643422)
(8.001087, 37.328614)
(9.001358, 39.214850)
(10.001466, 41.312997)
(11.001331, 42.886435)
(12.001397, 42.363187)
(13.001472, 42.781114)
(14.001347, 44.144667)
(15.001198, 45.823745)
(16.001236, 47.501800)
(17.001283, 47.606070)
(18.001261, 48.025022)
(19.001206, 45.824393)
(20.001223, 44.983532)
(21.001252, 40.998352)
(22.001423, 39.425334)
(23.001176, 37.328771)
(24.001420, 33.973193)
(25.001268, 30.092970)
(26.001463, 26.108759)
(27.001218, 28.413863)
(28.001230, 29.250054)
(29.001341, 30.299178)
(30.001173, 32.295801)
(31.001486, 40.259943)
(32.002149, 44.346740)
(33.001842, 52.536530)
(34.001411, 60.718155)
(35.001952, 69.420743)
(36.002487, 78.652719)
(37.001341, 83.263888)
(38.001069, 88.505846)
(39.001144, 93.327834)
(40.000705, 96.151593)
(41.001124, 96.267871)
(42.001541, 101.829704)
(43.001539, 101.082239)
(44.001973, 99.712891)
(45.001256, 98.670967)
(46.001123, 94.996125)
(47.001359, 91.011820)
(48.001765, 90.811897)
(49.001136, 92.595222)
(50.001141, 93.229018)
(51.001770, 94.689970)
(52.001233, 93.220395)
(53.001166, 93.220833)
(54.001135, 91.758391)
(55.001183, 89.131124)
(56.001194, 87.872654)
(57.001382, 86.932516)
(58.001091, 80.843905)
(59.001084, 70.147986)
(60.001120, 61.234550)
(61.001221, 52.635874)
(62.001565, 45.611372)
(63.001126, 38.902669)
(64.001252, 35.125508)
(65.001685, 32.502334)
(66.001346, 30.825786)
(67.001152, 29.882554)
(68.001215, 30.724089)
(69.001628, 34.603341)
(70.001405, 39.216613)
(71.001423, 43.514762)
(72.001166, 47.605668)
(73.001200, 51.589310)
(74.001307, 54.316670)
(75.001605, 56.101119)
(76.001255, 58.301897)
(77.001330, 59.872601)
(78.001190, 57.459949)
(79.001159, 52.846371)
(80.001201, 46.975133)
(81.001558, 45.509184)
(82.001423, 44.036728)
(83.001364, 41.941324)
(84.001093, 41.520414)
(85.001055, 42.988979)
(86.001471, 44.353164)
(87.001176, 47.186045)
(88.001287, 53.057327)
(89.001186, 62.390550)
(90.001089, 71.616872)
(91.001146, 77.279493)
(92.001194, 82.838473)
(93.001098, 89.548883)
(94.001205, 94.268415)
(95.001158, 98.146002)
(96.001215, 101.081090)
(97.001143, 103.071471)
(98.001113, 103.702634)
(99.001238, 100.241031)
(100.001253, 94.999933)
(101.001472, 90.066890)
(102.001293, 83.569350)
(103.001362, 74.132990)
(104.001192, 65.532351)
(105.001956, 57.040108)
(106.001236, 48.337198)
(107.000209, 39.426742)
(108.001335, 33.028537)
(109.001401, 28.205751)
(110.001349, 24.848945)
(111.001346, 20.344579)
(112.001452, 17.195788)
(113.001205, 17.089812)
(114.001791, 17.615891)
(115.001536, 18.244720)
(116.001575, 20.132179)
(117.001235, 17.930163)
(118.001328, 15.309001)
(119.001476, 14.260166)
(120.001443, 13.212204)

};
\addlegendentry{Slow (DL)}

\addplot [line width=1pt, ul, mark=x, mark size=2, mark options={fill=none}]
coordinates {
(1.000431, 20.549147)
(2.000212, 23.903721)
(3.000133, 26.944634)
(4.000422, 31.349953)
(5.000233, 34.390917)
(6.000492, 38.164338)
(7.000478, 34.498373)
(8.000183, 35.651197)
(9.000152, 36.384922)
(10.000537, 37.015754)
(11.000420, 37.640766)
(12.000436, 37.539556)
(13.000437, 38.378432)
(14.000217, 37.118382)
(15.001096, 37.741988)
(16.000242, 38.798915)
(17.000227, 39.742084)
(18.000373, 42.048591)
(19.001859, 43.936299)
(20.000431, 46.242634)
(21.000435, 47.504215)
(22.000505, 50.961422)
(23.000461, 51.801952)
(24.000415, 54.003972)
(25.000455, 55.686091)
(26.000502, 56.727341)
(27.000136, 56.832432)
(28.000220, 56.203934)
(29.000416, 52.428500)
(30.000489, 46.451342)
(31.000451, 42.047776)
(32.000408, 35.861368)
(33.000535, 34.706631)
(34.000510, 32.296066)
(35.000220, 31.142285)
(36.000458, 28.522157)
(37.000339, 26.318850)
(38.000097, 24.116745)
(39.000430, 24.746404)
(40.000158, 27.367967)
(41.000482, 29.779719)
(42.000439, 32.296453)
(43.000364, 31.352119)
(44.000317, 31.875932)
(45.000234, 30.933753)
(46.000225, 29.464954)
(47.000298, 28.627012)
(48.000230, 27.158657)
(49.000186, 26.528883)
(50.000134, 26.319754)
(51.000201, 25.061109)
(52.000328, 23.487917)
(53.000505, 22.754621)
(54.000211, 22.124464)
(55.000253, 20.551294)
(56.000458, 21.286106)
(57.000185, 20.027684)
(58.000459, 19.818065)
(59.000556, 17.826179)
(60.000198, 16.777329)
(61.000436, 16.777480)
(62.000516, 17.615945)
(63.000514, 17.930034)
(64.000195, 17.930809)
(65.000161, 18.560780)
(66.000463, 18.768866)
(67.000449, 20.132298)
(68.000449, 21.495078)
(69.000162, 24.221873)
(70.000429, 24.956018)
(71.000435, 25.480138)
(72.000607, 25.584937)
(73.000443, 26.423986)
(74.000168, 27.157918)
(75.000208, 28.311364)
(76.000521, 27.473392)
(77.000449, 27.892792)
(78.000469, 28.941944)
(79.000274, 29.045551)
(80.000189, 31.248042)
(81.000204, 31.561943)
(82.000181, 33.135602)
(83.000420, 32.716663)
(84.000174, 31.877860)
(85.000447, 31.142811)
(86.000437, 31.771474)
(87.000166, 30.303713)
(88.000171, 28.415932)
(89.000307, 26.214581)
(90.000435, 21.914361)
(91.000308, 19.293678)
(92.000436, 15.309049)
(93.000535, 12.791768)
(94.000477, 11.848213)
(95.000554, 9.751413)
(96.000487, 8.178651)
(97.000367, 6.920153)
(98.000283, 5.242699)
(99.000314, 5.872082)
(100.000547, 6.920749)
(101.000568, 7.654680)
(102.000528, 8.493497)
(103.000437, 11.010043)
(104.000350, 12.372904)
(105.000443, 14.470937)
(106.000496, 16.777930)
(107.000505, 18.141033)
(108.000549, 20.552255)
(109.000233, 23.592540)
(110.000262, 25.165428)
(111.000293, 27.158988)
(112.000473, 28.312316)
(113.000569, 28.206743)
(114.000525, 28.626400)
(115.000124, 28.625896)
(116.000199, 27.917453)
(117.000493, 26.239784)
(118.000468, 23.513977)
(119.000337, 18.900685)
(120.031568, 16.279130)
};
\addlegendentry{Slow (UL)}

\addplot [line width=1pt, dl_simulated, mark=x, mark size=2, mark options={fill=none}]
coordinates {
(1.001313, 2.516657)
(2.001496, 3.565583)
(3.001157, 4.404351)
(4.001087, 5.138075)
(5.001382, 5.347814)
(6.001048, 6.081914)
(7.001159, 6.606139)
(8.001539, 6.501336)
(9.001429, 6.501132)
(10.001297, 6.186429)
(11.001417, 6.710935)
(12.001428, 6.920337)
(13.001414, 8.073761)
(14.001491, 8.388839)
(15.001431, 8.703307)
(16.001401, 7.969207)
(17.001453, 7.444982)
(18.001237, 7.759274)
(19.001392, 7.654220)
(20.001129, 6.920273)
(21.000548, 6.081347)
(22.001088, 4.823018)
(23.001447, 3.343429)
(24.001474, 2.714171)
(25.001484, 2.581873)
(26.001395, 3.525459)
(27.001419, 3.944794)
(28.001147, 8.665389)
(29.001054, 10.762050)
(30.001194, 11.391154)
(31.001423, 12.230004)
(32.001095, 13.069157)
(33.001354, 13.709631)
(34.001421, 14.024308)
(35.001433, 14.366184)
(36.001085, 14.261358)
(37.001395, 14.366344)
(38.001097, 9.960612)
(39.001396, 7.654504)
(40.001516, 7.759142)
(41.001453, 7.339984)
(42.001444, 7.339732)
(43.001189, 6.396309)
(44.001546, 6.500781)
(45.001133, 5.766997)
(46.001085, 4.928237)
(47.001383, 4.403916)
(48.001582, 3.565048)
(49.001405, 3.250394)
(50.001537, 2.593614)
(51.001543, 2.046437)
(52.001587, 1.405543)
(53.001106, 1.226027)
(54.001182, 0.490557)
(55.001089, 0.490557)
(56.001426, 0.490557)
(57.001424, 0.490557)
(58.001448, 0.490557)
(59.000545, 0.490557)
(60.000949, 0.413596)
(61.000651, 0.331368)
(62.000624, 0.133361)
(63.000648, 0.312886)
(64.001150, 0.209663)
(65.001133, 0.209663)
(66.001096, 0.209663)
(67.001345, 0.314472)
(68.001350, 0.419400)
(69.000266, 0.524234)
(70.000535, 0.629024)
(71.000996, 0.943565)
(72.000330, 1.153327)
(73.000548, 1.048541)
(74.001199, 2.096992)
(75.001299, 2.830957)
(76.001075, 3.565176)
(77.000890, 4.194193)
(78.001010, 4.089265)
(79.001061, 6.292971)
(80.000768, 9.227211)
(81.001009, 12.911951)
(82.001160, 18.682050)
(83.000435, 25.492222)
(84.001041, 28.752881)
(85.000843, 29.754041)
(86.000345, 32.165440)
(87.001148, 34.991615)
(88.000990, 37.612392)
(89.000941, 36.353044)
(90.000976, 33.837994)
(91.001035, 30.573088)
(92.001288, 25.117330)
(93.000700, 18.831234)
(94.001289, 15.675971)
(95.000782, 14.360224)
(96.001135, 11.843927)
(97.001442, 9.018951)
(98.001084, 7.027575)
(99.001208, 6.187994)
(100.000931, 5.873778)
(101.001116, 5.244229)
(102.000710, 5.034685)
(103.001205, 5.034949)
(104.000974, 4.929742)
(105.001268, 6.922141)
(106.001312, 7.131291)
(107.001199, 7.025255)
(108.001139, 6.747545)
(109.001117, 6.537934)
(110.001583, 6.328170)
(111.001831, 6.223344)
(112.001562, 6.956934)
(113.001426, 8.214928)
(114.002846, 8.005121)
(115.001912, 6.390168)
(116.002319, 7.021457)
(117.002439, 7.754988)
(118.002473, 8.451868)
(119.002137, 9.290968)
(120.000944, 9.710290)

};
\addlegendentry{Fast (DL)}

\end{axis}

\end{tikzpicture}
    
    \caption{Evolution of the DL and UL throughput under ``Slow'' (limited speed and modest orientation changes) and ``Fast'' (complex maneuvers at higher speed) mobility conditions.}
    \label{fig:thr_gentle}
\end{figure}
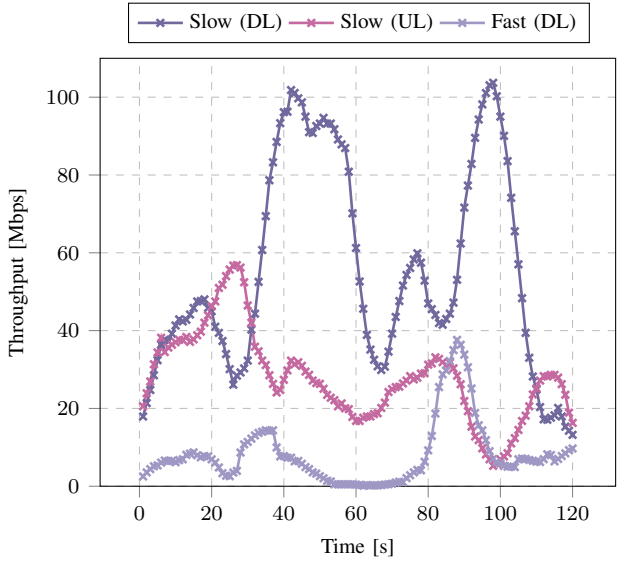

Then, in Fig.~\ref{fig:thr_gentle} we test the system under more dynamic conditions.
The \gls{uav} either flies slowly with limited speed, at about 2 m/s, and modest orientation changes (``Slow''), or executes more complex maneuvers at higher speed, changing attitude and pitch angles (``Fast'').
At high speed, we observe occasional link failures and throughput degradation, for example around 60 s. 
This is because the UAV dynamics may compromise the ability of the Starlink antenna to maintain precise alignment with the satellites.
Conversely, at slow speed, both DL and UL links remain fully usable, despite some fluctuations, even without manual or mechanical antenna pointing, which confirms the robust operation of the proposed~system.

\paragraph{Latency}

\begin{figure}
    \centering
    \input{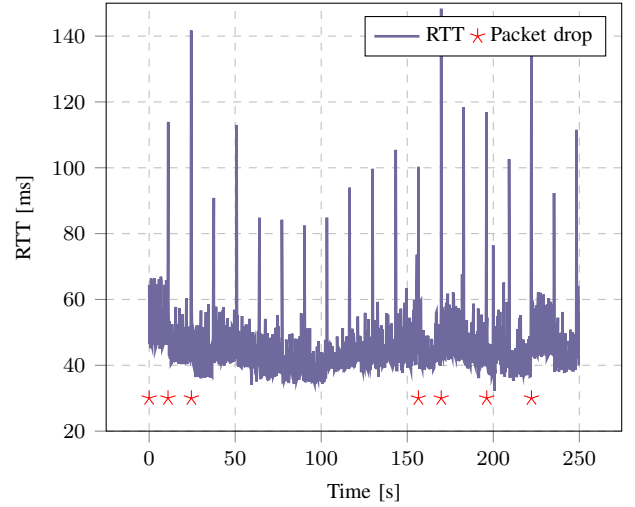}
    \caption{Evolution of the \gls{rtt}, with packet drop events.}
    \label{fig:latency}
\end{figure}

Latency in \gls{ntn} is predominately governed by the propagation delay due to the long distance between the ground and the satellites. In the case of Starlink, \gls{leo} constellations provide competitive \glspl{rtt} compared to other satellite solutions, often below 50 ms, but need frequent handovers as satellites move rapidly with respect to the Earth.

In Fig.~\ref{fig:latency}, we consider a static hovering scenario, and show the \gls{rtt} of the network for a single packet (ping),
measured from the time when data is generated at the ground UE to when it is received at the ground station via the UAV-satellite network, and in the reverse direction. Star marks indicate packet drop events. 
As expected, the \gls{rtt} is generally lower than 50 ms, though with periodic, high-magnitude spikes up to 140 ms at regular intervals of approximately 13 s. This systematic timing is indicative of Starlink-specific constellation management protocols that trigger handovers and require time-consuming signaling procedures, e.g., for path re-routing and re-initialization. 
We mainly attribute these spikes to either intra-satellite inter-beam handovers, where the \gls{ue} switches between different beams of the same satellite, or inter-satellite handovers, where the \gls{ue} associates with a different satellite.
Occasional packet drops are also recorded, suggesting that data loss may occur during~handover.


\begin{figure}
    \centering
\begin{tikzpicture}
\tikzset{every node/.style={font=\footnotesize}}

\definecolor{dl}{RGB}{0,119,136}
\definecolor{ul}{RGB}{237,117,100}

\definecolor{dl_simulated}{RGB}{159, 152, 196}
\definecolor{ul_simulated}{RGB}{232, 162, 205}

\begin{axis}[
tick align=outside,
tick pos=left,
xlabel={Time [min]},
xmajorgrids,
    ylabel={Battery voltage [V]},
    ymajorgrids,
    legend pos=north east,
    ymajorgrids=true,
    grid style=dashed,
    ytick={43, 44, 45, 46, 47, 48},
    ymin=42.3, ymax=48.7,
    tick label style={font=\footnotesize},
    ylabel style={font=\footnotesize},
    xlabel style={font=\footnotesize},
    legend style={at={(0.75,0.3)},legend cell align=left},
]

\addplot [line width=1pt, dl, dashed, mark size=2, mark options={fill=none}]
coordinates {
(0.000000, 48.400000)
(1.000000, 48.300000)
(2.000000, 48.280000)
(3.000000, 48.200000)
(4.000000, 48.100000)
(5.000000, 47.950000)
(6.000000, 47.880000)
(7.000000, 47.770000)
(8.000000, 47.670000)
(9.000000, 47.520000)
(10.000000, 47.370000)
(11.000000, 47.220000)
(12.000000, 47.100000)
(13.000000, 46.920000)
(14.000000, 46.700000)
(15.000000, 46.470000)
(16.000000, 46.240000)
(17.000000, 46.060000)
(18.000000, 45.710000)
(19.000000, 45.560000)
(20.000000, 45.430000)
(21.000000, 45.270000)
(22.000000, 45.100000)
(23.000000, 45.000000)
(24.000000, 44.880000)
(25.000000, 44.690000)
(26.000000, 44.500000)
(27.000000, 44.230000)
};
\addlegendentry{Without Starlink Mini antenna}

\addplot [line width=1pt, ul, mark=x, mark size=2, mark options={fill=none}]
coordinates {
(0.000000, 48.400000)
(1.000000, 48.125000)
(2.000000, 47.845000)
(3.000000, 47.680000)
(4.000000, 47.620000)
(5.000000, 47.540000)
(6.000000, 47.420000)
(7.000000, 47.365000)
(8.000000, 47.310000)
(9.000000, 47.245000)
(10.000000, 47.190000)
(11.000000, 47.095000)
(12.000000, 46.950000)
(13.000000, 46.800000)
(14.000000, 46.575000)
(15.000000, 46.270000)
(16.000000, 45.945000)
(17.000000, 45.520000)
(18.000000, 45.375000)
(19.000000, 45.205000)
(20.000000, 45.150000)
(21.000000, 44.990000)
(22.000000, 44.725000)
(23.000000, 44.555000)
(24.000000, 44.335000)
(25.000000, 44.070000)
(26.000000, 43.695000)
(27.000000, 43.305000)
};
\addlegendentry{With Starlink Mini antenna}

\end{axis}

\end{tikzpicture}
    
    \caption{Evolution of the battery voltage at the UAV, with and without the Starlink Mini antenna payload.}
    \label{fig:energy_consumption}
\end{figure}
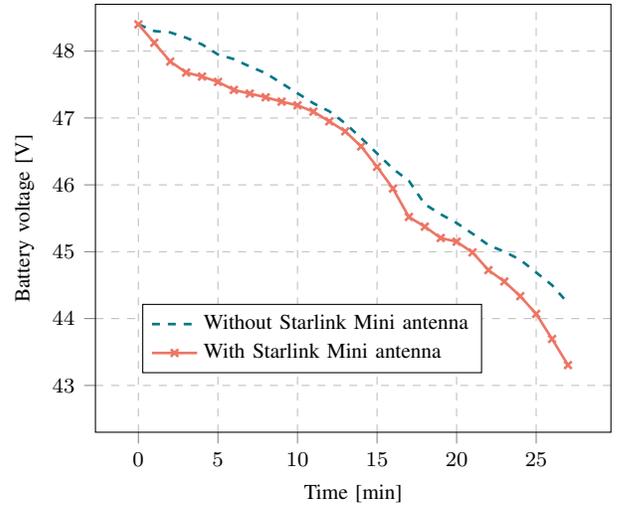

\paragraph{Energy consumption}
Finally, in Fig.~\ref{fig:energy_consumption} we evaluate the energy consumption at the UAV in terms of battery voltage levels both with and without the Starlink Mini antenna payload. 
We observe that the additional weight of the Starlink hardware has a minor impact on energy depletion. After 27 minutes of flight, the Starlink configuration showed a voltage drop of 0.9 V compared to the baseline. Given our operational range (48.5 V to 42 V representing 100\% to 30\% capacity), this corresponds to a 9\% reduction in remaining state of charge at the 27-minute mark. 
This result demonstrates that the proposed \gls{ppdr} system is not only able to effectively provide Internet connectivity, but also sustainable with respect to mission duration.

\section{Conclusions and Future Works}

\label{sec:conclusions}
In this paper we successfully demonstrated via field experiments near Catania (Italy) and ns-3 simulations a functional prototype in which a \gls{uav} drone, mounting a Starlink antenna terminal, rapidly provides Internet access and backhaul connectivity for \gls{ppdr} operations in otherwise unserved areas. 
The throughput analysis revealed an asymmetric behavior; DL capacity is primarily constrained by the local Wi-Fi link, with signal degradation at a distance of around 150-200 m; conversely, the UL capacity is bottlenecked by the satellite backhaul link.
Dynamic tests demonstrated that \gls{uav} velocity and maneuvering can, under certain circumstances, severely compromise the throughput, even though the system remains fully operational at moderate speeds, even with no mechanical antenna pointing.
The system supports a low RTT of around 40–50 ms, but systematic periodic latency spikes and occasional packet drop events were observed every 13 seconds, likely due to satellite handovers. Finally, the Starlink payload consumes little energy onboard the UAV, and the flight time is reduced by only 8\%.

As part of our future work, we will explore alternative terrestrial access technologies beyond Wi-Fi, such as 5G NR. We will also test a closed-loop \gls{ppdr} scenario where a UAV collects opportunistic 5G NR UL control signals to localize UEs, and communicates with the ground station for \gls{sar} support via the proposed UAV-satellite backhaul architecture.

\section*{Acknowledgements}
This work was partially supported by the European Union under the Italian National Recovery and Resilience Plan (NRRP) Mission 4, Component 2, Investment 1.3, CUP C93C22005250001, partnership on ``Telecommunications of the Future'' (PE00000001 -- program ``RESTART''). This work was also partially supported by the European Commission through the European Union’s Horizon Europe Research and Innovation Programme under the Marie Skłodowska-Curie-SE, Grant Agreement No. 101129618, UNITE.

\balance

\bibliographystyle{IEEEtran}
\bibliography{IEEEabrv,biblio}

\end{document}